\documentclass[a4paper,11pt]{article}
\pdfoutput=1
\usepackage{amssymb,amsmath,amsfonts,makeidx,placeins,multirow,tikz}
\usepackage{graphicx,rotate,subcaption,color,slashed,cite,caption,epstopdf,verbatim}
\usepackage{longtable,tabu}
\usepackage{array}
\usepackage[colorlinks=true,
            linkcolor=magenta,
            urlcolor=blue,
            citecolor=blue]{hyperref}

\numberwithin{equation}{section}
\numberwithin{figure}{section}
\numberwithin{table}{section}

\evensidemargin=\oddsidemargin
\begin{document}

\begin{flushright}
\begin{small}
TIFR/TH/18-44
\end{small}
\end{flushright}

\begin{center}

{\Large \bf Clockworked VEVs and  Neutrino Mass }\\
\vspace*{0.5cm} {\sf Avik
  Banerjee~$^{a,}$\footnote{avik.banerjeesinp@saha.ac.in}, ~Subhajit Ghosh~$^{b,}$\footnote{subhajit@theory.tifr.res.in}, ~Tirtha
    Sankar Ray~$^{c,}$\footnote{tirthasankar.ray@gmail.com}} \\
\vspace{10pt} {\small } $^{a)}$ {\em Saha Institute of Nuclear
    Physics, HBNI, 1/AF Bidhan Nagar, Kolkata 700064, India}
\\
\vspace{3pt} {\small } $^{b)}${\em Department of Theoretical Physics, Tata Institute of Fundamental Research, Homi Bhabha Road, Colaba, Mumbai 400005, India}
\\
\vspace{3pt} {\small } $^{c)}${\em Department of Physics and Centre
  for Theoretical Studies, Indian Institute of Technology Kharagpur,
  Kharagpur 721302, India} 
\normalsize
\end{center}


\begin{abstract}

In this paper we present an augmented version of the Abelian scalar clockwork model to generate geometrically suppressed vacuum expectation values (vev) of the pseudo Nambu-Goldstone bosons, that we call the \textit{clockworked} vevs. We briefly comment on generalization of the setup and possible 5D UV realizations.  We demonstrate how tiny neutrino mass can be generated by clockworking a weak scale vev.

\end{abstract}


\bigskip


\section{Introduction}
Generating stable hierarchical structures naturally has dominated the discourse  surrounding the  construction of models beyond the Standard Model. Varied scenarios have been proposed to explain the numerous hierarchies that has been observed in nature \cite{tHooft:1979rat,Kaul:2008cv,Bhattacharyya:2009gw,Csaki:2016kln}. A recent interesting proposal  is the \textit{Clockwork} mechanism \cite{Choi:2014rja,Choi:2015fiu,Kaplan:2015fuy,Giudice:2016yja,Lee:2017fin,Craig:2017ppp,Ben-Dayan:2017rvr} that utilises geometric suppression of couplings to explain the hierarchy of scale.

The Clockwork mechanism provides an elegant way to generate small couplings at low energy without resorting to  fine-tuning  in the UV. The scalar avatar of the  clockwork consists of a set of complex scalar fields located at different sites in theory space and each of which are charged under a site dependent global $\textrm{U(1)}$ symmetry that is spontaneously broken to a $\mathbb{Z}_2$ subgroup. In the ensuing theory of NGBs an additional clockwork potential explicitly  break the global $\textrm{U(1)}$ at each site to a single unbroken  $\textrm{U(1)}_{\textrm{CW}} $ by introducing nearest neighbour interactions between the sites.   The resulting spectrum contain  a true massless Nambu-Goldstone boson (NGB) while the others  acquire a tree level mass owing to the explicit breaking. The clockwork mechanism relies on the observation that the remaining massless NGB has a hierarchical distribution at different sites and utilises this hierarchy to generate geometric suppression in couplings. While the framework is easily  generalized to fermions and vector bosons and even gravitons \cite{Giudice:2017fmj,Hong:2017tel} the models in the literature are mostly confined to Abelian groups \cite{Craig:2017cda,Giudice:2017suc}, however, for exception see \cite{Ahmed:2016viu}. Interestingly clockwork potential has a simple 5D realization. It can be generated by deconstruction of the extra dimension on a discrete lattice \cite{Giudice:2016yja}. The clockwork mechanism has been applied to wide class of scenarios  that necessitates small couplings, for example, axion physics \cite{Kaplan:2015fuy,Giudice:2016yja,Coy:2017yex,Agrawal:2017cmd,Long:2018nsl}, dark matter scenarios \cite{Hambye:2016qkf,Kim:2017mtc,Kim:2018xsp,Goudelis:2018xqi}, inflation \cite{Choi:2014rja,Kehagias:2016kzt,Im:2017eju}, neutrino mass and flavour hierarchy \cite{Ibarra:2017tju,Patel:2017pct,Alonso:2018bcg}, relaxion models \cite{Choi:2015fiu,Flacke:2016szy,Davidi:2018sii} etc.    

In this paper we demonstrate the possibility to generate hierarchical vacuum expectation values (vev) using an augmented Abelian clockwork setup. A gap between the vev of a pNGB scalar and the scales involved in the Lagrangian can be obtained using the geometric suppression arising from clockwork mechanism, without introducing any substantial fine-tuning in the underlying theory. The clockwork gears have a hierarchic charge distribution under the remnant global $\textrm{U(1)}_{\textrm{CW}}$ symmetry which corresponds to a flat direction in the potential of the NGBs. We show that this remnant flat direction in the original clockwork potential can be lifted by addition of a soft breaking potential at any particular site such that, it ensures a vev for the pNGB at that site. The rest of the pNGBs at other sites then receive vevs hierarchically due to the clockwork mechanism. While the clockwork setup analogous to that in \cite{Giudice:2016yja} works in one direction, \textit{i.e.} the vev of the pNGBs monotonically decrease from one end of the clockwork to the other end, this can be extended  to frameworks where the clockwork can work along two directions. We illustrate the utility of the clockworked vev in explaining the smallness of neutrino mass. We show that a weak scale vev can be clockworked to generate the correct order of magnitude of neutrino mass. We also comment on the phenomenology of the clockwork gears in this context. A  possible  5D UV completion of this modified clockwork Lagrangian is sketched. 

In Section~\ref{review} we briefly review the basic clockwork setup. Generation of hierarchical vevs of the pNGBs in the extended clockwork scenario is discussed in Section~\ref{vev}. In Section~\ref{lepton} we apply this mechanism in a toy model to explain smallness of neutrino mass before drawing our conclusions in Section~\ref{conc}. 


\section{Review of  Scalar Clockwork}
\label{review}

The scalar clockwork setup comprises of $N+1$ sites in a theory space, each endowed with a global $\textrm{U(1)}$ symmetry spontaneously broken to its discrete subgroup $\mathbb{Z}_2$ at some scale $f$. In addition there are terms which simultaneously break this set of  $\textrm{U(1)}^{N+1}$ global symmetries explicitly to a single remnant $\textrm{U(1)}$. This setup consists of  a set of $N+1$ scalars $\Phi_j$ that is charged under the $j$-th $\textrm{U(1)}$ by a charge $q_j$, the resultant Lagrangian can be written as, 
\begin{equation}
\label{clockrev_1}
\mathcal{L}=\sum_{j=0}^{N}\left[\partial_\mu\Phi_j^\dagger\partial^\mu\Phi_j-\frac{\lambda}{8}(\Phi_j^\dagger\Phi_j-f^2)^2\right]+\frac{1}{2}\Lambda^{3-q}\sum_{j=0}^{N-1}\left(\Phi_j^\dagger \Phi^q_{j+1}+\textrm{h.c.}\right),
\end{equation}
where, the first two terms are invariant under the global $\textrm{U(1)}^{N+1}$, while the last term breaks the symmetry explicitly down to the remnant $\textrm{U(1)}_{\textrm{CW}}$. Note that, for $1 < q \le 3$, the explicit breaking term is renormalizable and soft (\textit{i.e.} $\Lambda\ll f$), whereas for $q>3$, the model becomes non-renormalizable and calls for further UV completion at scale $\Lambda\gg f$ \cite{Kaplan:2015fuy}. In this paper we will always assume the value of $q$ is within $(1,3)$\footnote{For $q=3$, the explicit breaking term, although independent of $\Lambda$, it may generate at scales lower than $f$ due to presence of some small coupling constant, which we have suppressed in Eq~\eqref{clockrev_1}.}.
The spontaneous breaking of the $\textrm{U(1)}^{N+1}$ gives rise to $N+1$ NGBs that can be parametrised by their non-linear representation,
\begin{equation}
\label{clockrev_2}
U_j=e^{i\frac{\pi_j}{f}}. 
\end{equation}
The radial modes which obtain masses $ \mathcal{O}(f)$, do not affect the NGB dynamics at low energy and can be ignored safely. While the  explicit breaking term  in Eq.~\eqref{clockrev_1} is not invariant under the independent $N+1$ shift symmetries ($\pi_j\rightarrow \pi_j+\alpha_j,~~\forall j$) of the NGBs, a remnant unbroken $\textrm{U(1)}_{\textrm{CW}}$ is left associated with the invariance under the transformation 
\begin{equation}
\label{clockrev_3}
\pi_j\rightarrow \pi_j+\frac{\alpha}{q^{j-1}},~~\forall j=0,...,N.
\end{equation}
The corresponding Abelian generator can be identified as,
\begin{equation}
\mathcal{T}=\sum_{j=0}^{N}{T_j \over q^j}~,
\end{equation}
where $ T_j $ is the generator of j-th site. Clearly, out of total $(N+1)$ NGBs, only one remains massless while the other modes develop non-trivial tree level masses proportional to the strength of explicit breaking term. The corresponding  pNGB potential can be calculated using Eq.~\eqref{clockrev_1} as
\begin{equation}
\label{clockrev_4}
V_{\pi}=-\frac{1}{2} f^{q+1}\Lambda^{3-q}\sum_{j=0}^{N-1}\left(U_j^\dagger U^q_{j+1}+\textrm{h.c.}\right)=-f^{q+1}\Lambda^{3-q}\sum_{j=0}^{N-1}\cos\left(\pi_j-q\pi_{j+1} \over f\right).
\end{equation} 
It is easy to identify the flat direction corresponding to a true Goldstone boson in the vacuum configuration of the above potential. The mass matrix can be calculated as 
\begin{equation}
\label{clockrev_5}
M_\pi^2=f^{q-1}\Lambda^{3-q}\left(\begin{array}{cccccc}
 1& -q &  0 & ...& 0 & 0 \\ 
-q & q^2+1 & -q & ...& 0 & 0 \\ 
0 & -q & q^2+1 & ...& 0  &  0\\ 
\vdots & \vdots & \vdots & \ddots & \vdots &\vdots \\ 
0 & 0 & 0 &  & q^2+1& -q\\
0 & 0 & 0 &...  & -q & q^2
\end{array} \right).
\end{equation}
The tridiagonal symmetric mass matrix shown above can be diagonalized by an orthogonal rotation ($\pi_j=O_{jk}a_k$) to the eigenbasis yielding one massless mode and $N$ massive modes as \cite{Kaplan:2015fuy,Giudice:2016yja}
\begin{equation}
\label{clockrev_6}
m_0^2=0,~~m_k^2=\lambda_k f^{q-1}\Lambda^{3-q},~~\textrm{where,}~~\lambda_k\equiv q^2+1-2q\cos\frac{k\pi}{N+1},~k=1,...,N~.
\end{equation}
The diagonalizing matrix is given by
\begin{eqnarray}
\label{clockrev_7}
O_{j0}=\frac{1}{q^j}\sqrt{\frac{q^2-1}{q^2-q^{-2N}}},~~
O_{jk}=\sqrt{\frac{2}{(N+1)\lambda_k}}\left[q\sin\frac{jk\pi}{N+1}-\sin\frac{(j+1)k\pi}{N+1}\right],
\end{eqnarray}
where $j=0,...,N$ and $k=1,...,N$. This shows that the massless eigenstate ($a_0=O_{j0}\pi_j$) is hierarchically localized at the different sites with a weight factor $1/q^j$, which for large values of $j$ can give rise to an exponential suppression. This observation  leads to the clockwork effect.


\section{Clockworked VEVs}
\label{vev}

In this section we discuss the augmented clockwork mechanism to produce geometrically suppressed vevs. We will present the modified clockwork potential required to facilitate this and demonstrate how such a setup can naturally generate a suppressed vev without fine-tuning in the underlying  theory.
 
 

The clockwork potential is invariant under an unbroken global  $\textrm{U(1)}$ and spontaneous breaking of that symmetry gives rise to a flat direction in the Goldstone potential. However,  the resulting  NGBs arising in Eq.~\eqref{clockrev_4} posses a discrete $\mathbb{Z}_2$ symmetry and can not receive vevs. To generate a hierarchical vev structure, the primary requirement is to simultaneously  lift that remnant flat direction and explicitly break the $\mathbb{Z}_2$ symmetry. We introduce a potential that explicitly break the  residual $\textrm{U(1)}_{\textrm{CW}}$ as well as the corresponding discrete symmetry while  minimising the NGBs at non zero field configurations. Here we consider the breaking potential at a particular site resulting in a vev for the pNGB at that site. At this stage the original clockwork potential comes into play by communicating the vev generated at a particular site to all other sites in a hierarchic manner. 

\begin{figure}[t]
	\centering
	\begin{subfigure}[t]{0.45\textwidth}
		\centering
		\includegraphics[width=\linewidth]{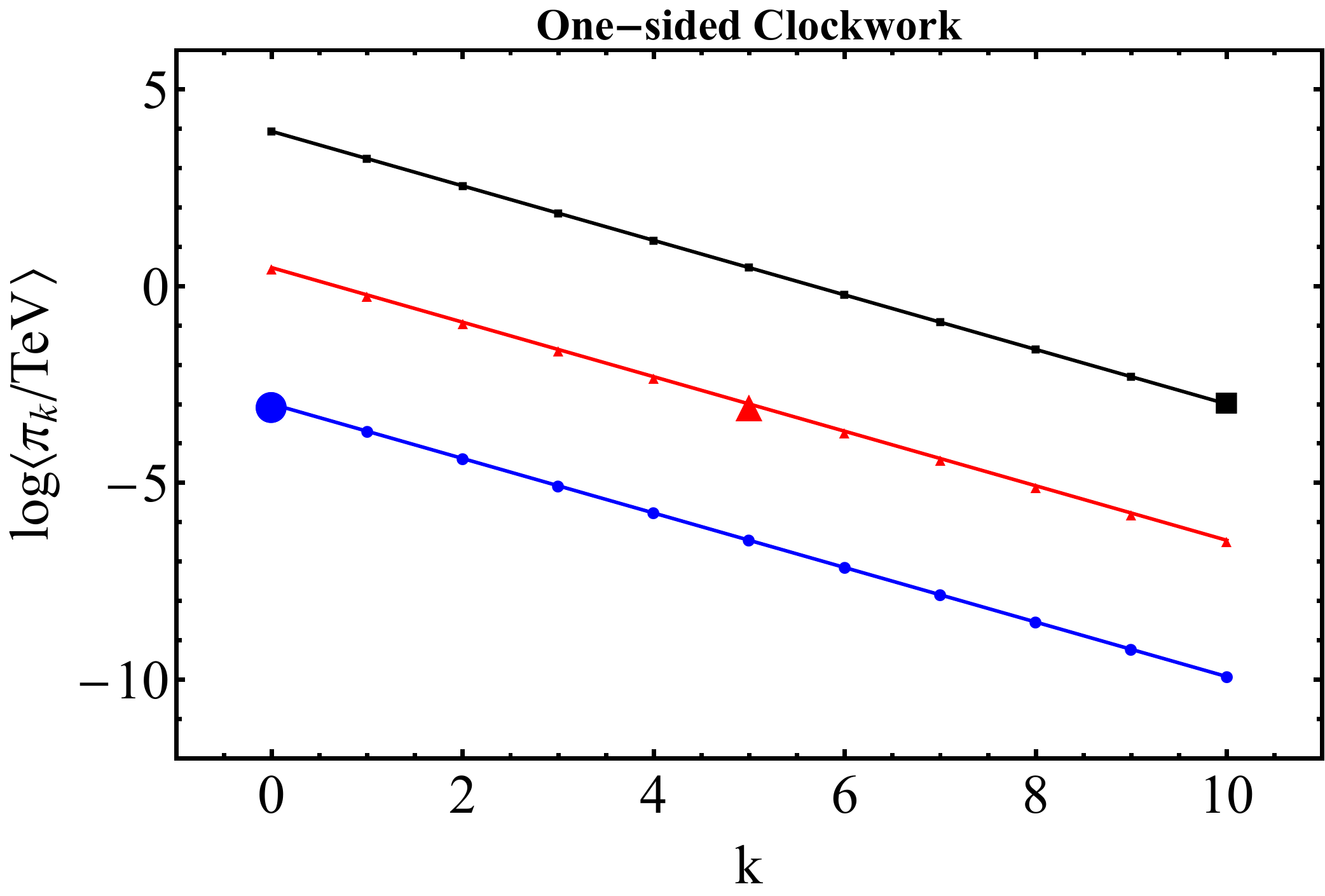}
	\end{subfigure}
	~
	\begin{subfigure}[t]{0.445\textwidth}
		\centering
		\includegraphics[width=\linewidth]{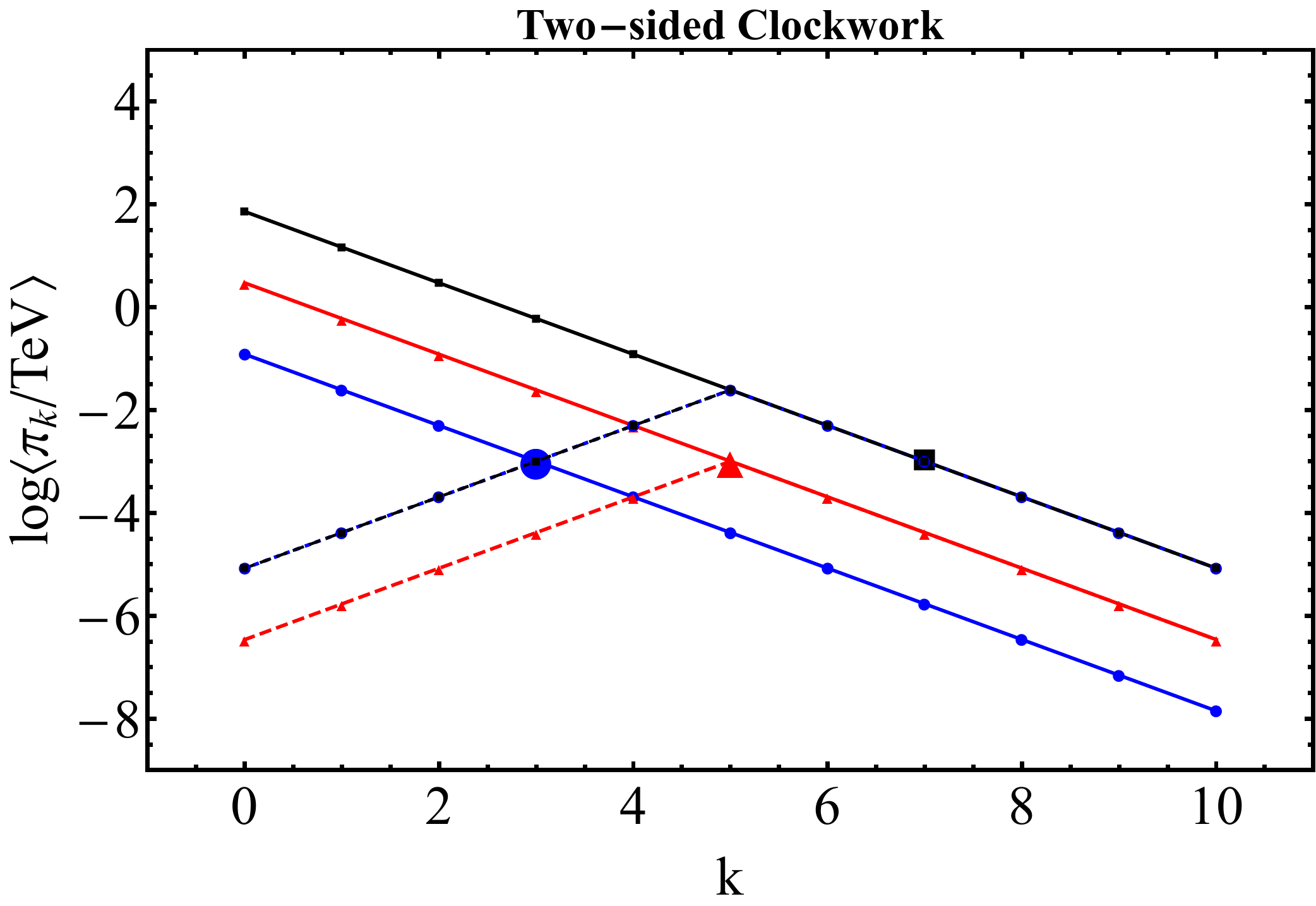}
	\end{subfigure}
	\caption{\small\it The left panel shows an illustrative example of the variation of $\log\langle\pi_k/\textrm{TeV}\rangle$ as a function of the site $k$ for a one-sided clockwork setup with total $11$ sites (i.e. $N=10$). Three different lines correspond to the soft breaking potential $V_{\textrm{soft}}$, added at sites $k=0$ (blue), $k=5$ (red) and $k=10$ (black) respectively. The bigger bullets also indicate the site of the breaking term. Notice that, for all three lines, the largest vev is located at site $k=0$ and smallest at $k=10$, irrespective of the site of breaking. The \textit{two-sided} clockwork setup is shown in the right panel where pivot point is at the middle site $k=5$, while the soft breaking terms are added at $k=3$ (blue), $k=5$ (red) and $k=7$(black). Three solid lines correspond to the one-sided clockwork while the dashed lines shows the \textit{two-sided} case. Notice that, for all three lines, the largest vev is located at site $k=5$ and smallest at $k=0$ and $k=10$, irrespective of the site of breaking. In this case, soft breaking at $k=3$ and $k=7$ gives rise to same structure of vev hierarchy.}
	\label{vev_fig}
\end{figure}

Consider the following soft breaking terms ($\mu_1,\mu_2\ll f$) in the NGB potential in addition to the standard clockwork potential given in Eq.~\eqref{clockrev_4},
\begin{equation}
\label{oneside_1}
V_{\textrm{soft}}=-\frac{\mu_1^{2}f^{2}}{4}\left(U_k+\textrm{h.c}\right)^2+\frac{\mu_2^{3}f}{2}\left(iU_k+\textrm{h.c}\right)=-\mu_1^{2}f^{2}\left(\cos {\pi_k \over f} \right)^2 - \mu_2^{3}f\sin{\pi_k \over f},
\end{equation}  
where $k$ refers to any particular site from $0$ to $N$. A possible 5D UV realization of the model is discussed in Appendix \ref{deconst}. The breaking terms above do not necessarily introduce new scales in the clockwork setup as $\mu_1,\mu_2$ can be of the order of $\Lambda$ without any loss of generality. Note that both the term breaks the remnant $ \textrm{U(1)}_{\textrm{CW}} $ and the last term breaks the $\mathbb{Z}_2$ symmetry explicitly. Let us assume the soft terms are added at the zeroth site ($k=0$). The minimization condition for the total potential ($V=V_\pi+V_{\textrm{soft}}$) then yields
\begin{equation}
\label{oneside_2}
\sin\left(2\langle\pi_0\rangle \over f\right) \left(\cos\left(\langle\pi_0\rangle \over f\right)\right)^{-1} = {\mu_2^3 \over f\mu_1^2},~~\langle\pi_1\rangle=\frac{\langle\pi_0\rangle}{q},~\langle\pi_2\rangle=\frac{\langle\pi_0\rangle}{q^2},~...,~\langle\pi_N\rangle=\frac{\langle\pi_0\rangle}{q^N}.
\end{equation}
Expanding the trigonometric functions, at the leading order  we get the vev of the pNGBs as
\begin{equation}
\label{oneside_3}
 \langle\pi_0\rangle = {\mu_2^3 \over 2\mu_1^2},~~\langle\pi_1\rangle=\frac{\langle\pi_0\rangle}{q},~\langle\pi_2\rangle=\frac{\langle\pi_0\rangle}{q^2},~...,~\langle\pi_N\rangle=\frac{\langle\pi_0\rangle}{q^N}.
 \end{equation} 
Eq.~\eqref{oneside_3} clearly shows the clockwork setup develops a hierarchical vev structure with the minimum vev arising at the farthest end from the soft breaking site. If instead of the $0^{\textrm{th}}$ site, the potential of Eq.~\eqref{oneside_1} is added at some generic $k^{\textrm{th}}$ site, the same kind of hierarchical vev structure will appear with $\langle\pi_k\rangle=\mu_2^3/2\mu_1^2$. However, the vevs of the pNGBs at sites less than $k$ will be larger than $\langle\pi_k\rangle$ while that for pNGBs at sites greater than $k$ will be smaller. We call this structure as \textit{`one-sided'} clockwork as illustrated pictorially in the left panel of Fig.~\ref{vev_fig}. 

Interestingly the clockwork vevs do not spoil the original structure of the clockwork mass matrix. However, due to the additional soft breaking term, the zero mode of the original clockwork setup receives a mass. The mass matrix can be written as
\begin{equation}
\label{oneside_5}
M^2_\pi\simeq M^2_{(0)}+ \mu_1^2M^2_{(1)}~,
\end{equation} 
where $M^2_{(0)}$ is the mass matrix as given in Eq.~\eqref{clockrev_5},  and $M^2_{(1)}=\textrm{diag}\left(2,0,0,...,0\right)$ denotes the contribution due to the soft breaking. 
For simplicity of calculation, assuming  $\mu_1,\mu_2\ll\Lambda$  the eigenvalues and eigenstates can be expressed at the leading order as
\begin{equation}
\label{oneside_6}
m^2_{j}\simeq m^{2}_{j(0)}+\mu_1^2 m^{2}_{j(1)},~~\textrm{and},~~a_j\simeq a^{(0)}_j+\mu_1^2 a^{(1)}_j,
\end{equation}
where we have neglected corrections of the order $\mathcal{O}\left( {\mu_1^4\over (f^{q-1}\Lambda^{3-q})^2}\right).$
Using Eqs.~\eqref{clockrev_6},~\eqref{clockrev_7}  in the above expression we obtain
\begin{equation}
\label{oneside_8}
m^2_0\simeq 2\mu_1^2O_{00}^2,~~m^2_k\simeq\lambda_k f^{q-1}\Lambda^{3-q}+2\mu_1^2O_{k0}^2~.
\end{equation}
Note that the exact parametric dependences of the masses will be different if the soft breaking terms are added at sites other than the zeroth site and can be calculated analogously. To find the eigenstates, we assume that $a_j^{(1)}$ can also be written as a linear combination of $a_j^{(0)}$, and at the same order in the expansion can be expressed as 
\begin{eqnarray}
\label{oneside_10}
a_j\simeq a_j^{(0)}+2\mu_1^2f^{1-q}\Lambda^{q-3}\sum_{k\ne j}\frac{O_{k0}O_{j0}}{\lambda_j-\lambda_k}a^{(0)}_k=O_{ij}\pi_i+2\mu_1^2f^{1-q}\Lambda^{q-3}\sum_{k\ne j}\frac{O_{k0}O_{j0}}{\lambda_j-\lambda_k}O_{lk}\pi_l~.
\end{eqnarray} 
Here we note in the passing that  the potential  given in Eq.~\eqref{oneside_1} is not unique  in generating a hierarchical one-sided clockwork vev \footnote{For generalization of the clockwork mechanism including next-to-nearest neighbour interactions and for supersymmetrized version of clockwork, see \cite{Ben-Dayan:2017rvr}.}. The generalization includes higher powers of $U_k$ in the breaking potential as,
\begin{equation}
\label{oneside_13}
V_{\textrm{soft}}=-\frac{\mu_1^{4-2p}f^{2p}}{4}\left(U_k^{p}+\textrm{h.c}\right)^2+\frac{\mu_2^{4-r}f^{r}}{2}\left(iU_k^{r}+\textrm{h.c}\right),
\end{equation}  
where we assume $1\le (p,r)\le 4$ for renormalizability. One can  also generalise the breaking term to accommodate nearest neighbour interactions analogous to the standard clockwork potential in Eq.~\eqref{clockrev_4}
\begin{eqnarray}
\label{oneside_11}
V_{\textrm{soft}}&=-\frac{\mu_1^{4-2p-2p^\prime}f^{2p+2p^\prime}}{4}\left(U_k^{p} U_{k+1}^{\dagger p^\prime}+\textrm{h.c}\right)^2+\frac{\mu_2^{4-r-r^\prime}f^{r+r^\prime}}{2}\left(iU_k^{r}U_{k+1}^{\dagger r^\prime}+\textrm{h.c}\right).
\end{eqnarray}  
Clearly, in the limit $r^\prime=p^\prime=0$, we get back Eq.~\eqref{oneside_13}. While the vev of $\pi_k$ changes due to the minimization of this potential, the clockworked vev structure remains similar to the previous case.   
It can be shown that, even further generalization to next-to-nearest neighbour breaking terms (e.g. $U^p_kU^{p^\prime}_{k+1}U^{\dagger p^{\prime\prime}}_{k+2}+\textrm{h.c.}$) can also gives rise to similar scenario. In principle any combination of fields that breaks the remnant $\textrm{U(1)}_{\textrm{CW}}$ can be used to generate vev.  While the calculations are straightforward and follow analogous to the discussion in this section, we will not  explore these possibilities further in this paper.
 
Another different possibility is a clockwork effect that can propagate in two directions along the sites. This is a simple generalisation of the clockwork  framework presented in  Section \ref{review}. This can be viewed as gluing together of two clockwork Lagrangian with scalar fields having opposite charge orientations. In the resultant chain the location where the potential crosses over from one clockwork to the other is a pivot site where the suppression  factor undergoes an inflection. We modify and extend the clockwork potential to the form,
\begin{eqnarray}
\label{twoside_1}
\nonumber
V_{\pi}&=-\frac{1}{2} f^{q+1}\Lambda^{3-q}\left[{1\over q^2}\sum_{j=0}^{k-1}\left(U_j^{\dagger q} U_{j+1}+\textrm{h.c.}\right)+\sum_{j=k}^{N-1}\left(U_j^\dagger U^q_{j+1}+\textrm{h.c.}\right)\right],\\
\nonumber
\\
&=-f^{q+1}\Lambda^{3-q}\left[{1\over q^2}\sum_{j=0}^{k-1}\cos\left(q\pi_j-\pi_{j+1} \over f\right)+\sum_{j=k-1}^{N-1}\cos\left(\pi_j-q\pi_{j+1} \over f\right)\right].
\end{eqnarray} 
 In analogy to Eq.~\eqref{clockrev_1}, in this case the $\textrm{U(1)}_{\textrm{CW}}$ charge for the field $\Phi_k$ is $1$ whereas that for fields $\Phi_{k\pm j}$ is $1/q^{j}$.  Here, $k$ represent the pivot site \footnote{Here, the factor $1/q^2$ in front of the first term in the potential is kept for matching this scenario with the 5D realization discussed in Section \ref{deconst}. It has, however, no significant effect in the discussion below.}. For simplicity, let us assume $N$ is even and the pivot point $k=N/2$.  This would lead to a \textit{`two-sided'} clockwork. One can further utilise this two-sided clockwork to generate progressively suppressed vev   transmitted along the two directions around the pivot site.  As indicated in the previous section, a soft breaking potential of the form given in  Eq.~\eqref{oneside_1} at the pivot site  $k=N/2$ leads to a vev structure of the pNGBs as
\begin{equation}
\label{twoside_2}
 \langle\pi_0\rangle =\frac{\langle\pi_{k}\rangle}{q^k},~~...~~\langle\pi_{k-1}\rangle=\frac{\langle\pi_{k}\rangle}{q},~~\langle\pi_{k}\rangle={\mu_2^3 \over 2\mu_1^2},~~\langle\pi_{k+1}\rangle=\frac{\langle\pi_{k}\rangle}{q},~~...~~\langle\pi_N\rangle=\frac{\langle\pi_{k}\rangle}{q^k}~.
 \end{equation} 
The right panel of Fig.~\ref{vev_fig} illustrates the two-sided clockwork vev generation mechanism. Note that the hierarchic vev structure is symmetric with respect to the pivot site \footnote{Asymmetric structure can be constructed with odd values of $N$ or with different values of $q$ on different sides of the pivot point.}. 
Mass matrix of this setup can be extracted using the potential in Eqs.~\eqref{twoside_1} and \eqref{oneside_1}, resembling  that of two oppositely oriented clockworks joined end to end. Similar to the one-sided case, in the limit $\mu_1^2\to 0$, it leads to one zero mode in the spectrum which is localized in a \textit{two-sided} hierarchic manner.  When the soft breaking terms are turned on (\textit{i.e.} $\mu_1^2\ne 0$), the massless NGB mode will receive a small mass correction proportional to the strength of the explicit breaking similar to the form given in Eq.~\ref{oneside_6}.
 

\section{A Toy Model for Neutrino Mass}
\label{lepton}

\begin{table}[t]
	\begin{longtable}{ccccccccc}
		\hline 
		\rule[-2ex]{0pt}{5.5ex} Fields &  & $\textrm{SU(2)}_L$ & & $\textrm{U(1)}_{Y}$ & $\mathbb{Z}_2^{(j)}$ \\ 
		\hline  
		\hline
		\rule[-2ex]{0pt}{5.5ex} $H$ & & $0$ & & $\frac{1}{2}$ & $+$\\
		\rule[-2ex]{0pt}{5.5ex} $\pi_{j}$ & & 1 & & $0$ & $-$\\ 
		\rule[-2ex]{0pt}{5.5ex} $l_{L}$ & & 2 & & $-\frac{1}{2}$ & $+$\\ 
		\rule[-2ex]{0pt}{5.5ex} $\nu_{R}$ & & 1 & & $0$ & $-$\\
		
		\hline 
		
		\caption{Particle content of the toy model} 
		\label{flav_tab1}
	\end{longtable}
\end{table}

\begin{figure}[t]
	\centering
	\includegraphics[width=0.45\textwidth]{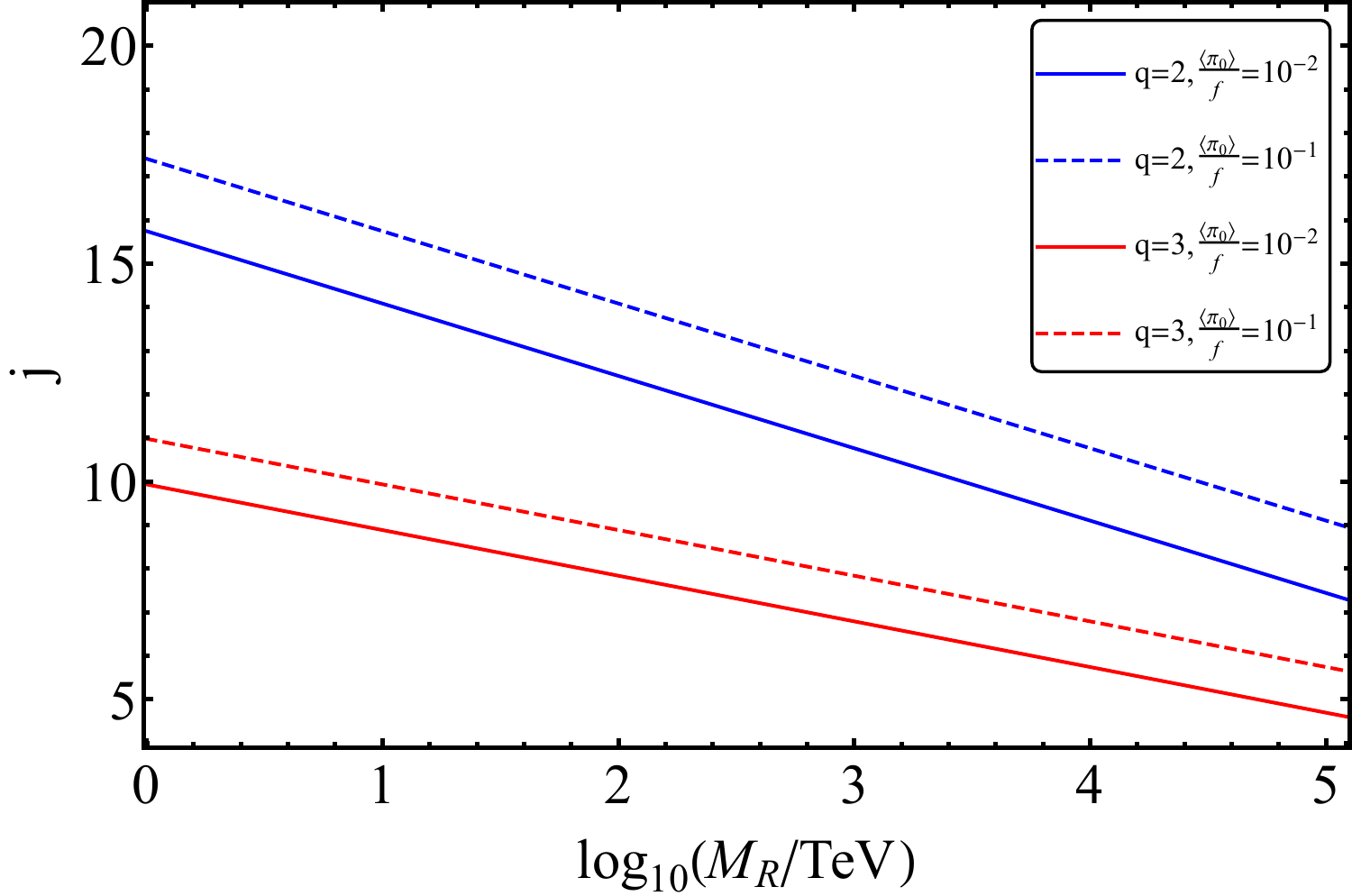}
	\caption{\small\it The figure shows variation of the site
		number at which the right handed neutrino should couple to the one-sided clcokwork setup as a function of its Majorana mass. We present in the solid lines the variation assuming $\langle\pi_0\rangle/f\sim10^{-2}$ and in dashed lines assuming when $\langle\pi_0\rangle/f\sim10^{-1}$ for both $q=2$ (blue) and $q=3$ (red).}
	\label{flav_fig1}
\end{figure}

In this section we present a rough sketch for utilising the clockworked vev to generate tiny neutrino masses. Let us assume the right handed neutrino possesses a charge under the $\mathbb{Z}_2$ of the $j^{\textrm{th}}$ site of the clockwork chain, denoted by $\mathbb{Z}_2^{(j)}$. The particle content of the model is shown in Table~\ref{flav_tab1}. We introduce a higher dimensional operator which generates a Dirac-like mass term for the neutrino as
\begin{equation}
\label{flav_1}
\mathcal{L}_{\nu}=y^\nu\left(\frac{\pi_j}{f}\right)\bar{l}_{L}H^c\nu_{R}+\textrm{h.c.},
\end{equation}
where $y^\nu$ denotes some effective coupling strength. We assume that this operator is dynamically generated at scales where the $\mathbb{Z}_2^{(j)}$ is an exact symmetry (\textit{i.e.} between scales $f$ and $\Lambda$), however, we remain completely agnostic about the origin of such operator, which depends on the specific UV completion.  The Dirac mass of the neutrino is estimated as
\begin{equation}
\label{flav_3}
m_\nu={y^\nu v\over\sqrt{2}}{\langle\pi_j\rangle\over f}\simeq{y^\nu v\over\sqrt{2}}{\langle\pi_0\rangle\over f}{1\over q^j}~,
\end{equation}	
Here, we utilise the one sided clockwork vev structure as given in Eq.~\eqref{oneside_3} and $v/\sqrt{2}$ is the vev of the Higgs. Assuming $y^\nu\sim\mathcal{O}(1)$, $q=3$ and $\langle\pi_0\rangle/f\sim\mathcal{O}(10^{-1})$, we find that the right handed neutrino should couple to the site given by $j\simeq24$ of a clockwork chain, in order to reproduce the neutrino mass $\sim0.1~\textrm{eV}$. Thus a weak scale vev upon clockworking gives rise to the scale of neutrino mass. In effect the necessity of a small Yukawa coupling is traded off by the geometric suppression arising by the choice of the site at which right handed neutrino couples to the clockwork chain. Further if the right handed neutrinos have some Majorana mass, `TeV scale seesaw' can be achieved using the clockwork vevs. The seesaw mass formula for the neutrino is given as
\begin{equation}
\label{flav_5}
m_\nu\simeq{m_D^2\over M_R}=\left({y^\nu v\over\sqrt{2}}{\langle\pi_0\rangle\over f}{1\over q^j}\right)^2{1\over M_R}~,
\end{equation} 
where $M_R$ denotes the Majorana mass of the right handed neutrino and $m_D$ corresponds to the Dirac mass, arising from Eq.~\eqref{flav_1}. Assuming $M_ R\sim\mathcal{O}(\textrm{TeV})$, we get $j\simeq 10$ to obtain $m_\nu\sim0.1~\textrm{eV}$, for variation of $j$ with the Majorana mass of the right handed neutrino ($M_R$), see Fig.~\ref{flav_fig1}. Interestingly one can envisage the possibility of generating an anarchic structure of neutrino mass matrix in this model by coupling all the generations of right handed neutrinos to the same site in the clockwork chain \cite{Hall:1999sn}. Extending the clockwork setup to other types of seesaw mechanisms, analogous to \cite{Gehrlein:2018sdk}, are also worth investigating.

Admittedly, below the scale $\Lambda$, the clockwork potential (Eq.~\eqref{clockrev_4}) breaks the $\mathbb{Z}_2^{(j)}$ symmetry explicitly and leads to a mass mixing between nearest neighbour clockwork states. This breaking 
will lead to interactions of all the clockwork quantum states with the neutrino and the Higgs fields. The interaction term, after inverting Eq.~\eqref{oneside_10} and keeping only the leading term, can be written as
\begin{equation}
\label{flav_6}
\mathcal{L}_{\nu}\simeq y^\nu\left[O_{ji}\left(\delta_{ik}-\mathcal{O}(\mu_1^2f^{1-q}\Lambda^{q-3})\right)\right]{a_k\over f}\bar{l}_{L}H^c\nu_{R}+\textrm{h.c.}
\end{equation}
We assume that $f$ is around TeV scale and all the clockwork gears except the lightest one are heavier than the Higgs, while the mass of the lightest eigenstate ($a_0$) depends on the scale $\mu_1$. Assuming $\mu_1\sim100~\textrm{GeV}$ and $q=3,~ y=\mathcal{O}(1)$, we find the kinematically allowed Higgs invisible decay width to $\nu\bar{\nu}a_0$ is very small ($\Gamma_{h\to\nu\bar{\nu}a_0}\sim 10^{-18}~\textrm{GeV}$) and virtually unconstrained from the current LHC data. The collider phenomenology of the heavy neutrinos also depend on their couplings with the weak gauge bosons, which are proportional to the mixing between the light and heavy states ($\sim m_D/M_R$) \cite{Han:2006ip,delAguila:2007qnc,Kersten:2007vk,delAguila:2008cj}.  Although, TeV-scale seesaw as presented here, provides larger mixing compared to the vanilla type-I seesaw, significant signatures at the LHC would require additional non-standard interactions of the singlet neutrinos, e.g. with a $Z^\prime$ gauge boson in a gauged $B-L$ extension \cite{Perez:2009mu}.
 
Note that the operator in Eq.~\eqref{flav_1} results in a hard breaking of the clockwork shift symmetry and therefore may lead to significant radiative corrections to the clockwork potential. Without going into a detail analysis of Coleman-Weinberg potential, we resort to an order of magnitude estimate using dimensional analysis that shows the divergent loop corrections generated at one-loop and two-loops respectively, as given below
\begin{equation}
\Delta\mathcal{L}\sim -(y^\nu)^2{M^2v^2\over 16\pi^2}\left({\pi_j\over f}\right)^2-(y^\nu)^2{M^4\over (16\pi^2)^2}\left({\pi_j\over f}\right)^2. 
\end{equation}
Assuming, the cut-off $M\lesssim 4\pi f$, one can check that the radiative corrections do not spoil the hierarchy of the clockworked vevs, however it generates a small correction to the magnitude of the vevs. The contributions coming from even higher loop orders can be neglected in comparison to the above, due to further loop suppressions.

As an aside, we mention that the masses of the charged leptons can also be generated in similar ways. We will consider that the right handed component of the charged leptons are charged under  a family dependent $\mathbb{Z}_2^{(j)}.$ Assuming the same ranges for the different relevant parameters as for the neutrino case, we determine the ratio between the sites at which $e_R$, $\mu_R$ and $\tau_R$ should couple to the clockwork setup to produce the correct order of masses are
\begin{equation}
\label{flav_7}
{j_e\over j_\mu}\simeq 2,~~{j_\mu\over j_\tau}\simeq 2~.
\end{equation}
However, the precision measurements of the anomalous magnetic moment $(g-2)$ of the charged leptons can provide strong constraints on this setup. Below the electroweak symmetry breaking scale, an effective Yukawa coupling between the charged leptons and all the clockwork gears ($va_k\bar{l}l/f $) may lead to significant contributions to the $(g-2)$ of leptons via one-loop diagrams \cite{Jackiw:1972jz,Jegerlehner:2009ry}. The one-loop contributions coming from these clockwork states are found to be positive definite, while latest precision measurements of the $(g-2)$ of electrons show a negative deviation from the standard model value \cite{PhysRevLett.100.120801,Hanneke:2010au}, in tension with the prediction of the present framework. On the other hand, it may provide a handle to address the persistent positive deviation of $(g-2)$ for muons \cite{Bennett:2006fi,Blum:2018mom}.


\section{Conclusions}
\label{conc}

In this paper we demonstrate that a simple extension of the clockwork mechanism can be employed  to successfully generate hierarchical structure of vevs for the pNGB scalars. We show that  crucial to generating clockworked vev is introduction of simultaneous lifting of the residual flat direction and explicit breaking of discrete symmetries beyond the standard clockwork set up. This can be done by adding appropriate terms  at a particular site in the  clockwork chain. The additional term in the clockwork potential is found not to be unique, rather we illustrate a class of scenarios where this can be achieved. We also show that both the original clockwork as well as the augmented version involving pNGB vevs,  can have a \textit{two-sided} hierarchy structure. However, explaining the intrinsic hierarchy between the two relevant scales of the clockwork mechanism, namely $f$ and $\Lambda$ as well as constructing proper UV completion of the clockworked vev setup is challenging and worth further studies. 

Clockworked vevs can be used for a wide range of applications. We discuss its utility  to generate small neutrino mass without fine-tuning the underlying theory of electroweak symmetry breaking. For Dirac neutrino masses, the right handed neutrinos has to couple to a `flavour depended' site in the clockwork chain. For the seesaw models the clockworked vevs lower the seesaw-scale to TeV, while generating the correct order of magnitude of the neutrino mass. One can utilise an analogous setup to generate hierarchical masses for the other standard model fermions. It will indeed be interesting to see how the mixing between lepton flavours as encoded by the PMNS matrix can originate from this framework.


\begin{small}
\section*{Acknowledgments}

We thank Gautam Bhattacharyya, Emilian Dudas, Rick S. Gupta and Tuhin S. Roy for useful discussions and comments. AB and SG acknowledge support from Department of Atomic Energy, Government of India. AB also likes to acknowledge support by the (Indo-French) IFCPAR/CEFIPRA Project No. 5404-2 and hospitality of LPT, Orsay and \'Ecole Polytechnique, Palaiseau during the final stages of the work. TSR is partially supported by the Department of Science and Technology, Government of India, under the Grant Agreement number IFA13-PH-74 (INSPIRE Faculty Award).

\end{small}


\appendix

\section{UV completion from 5D deconstruction}
\label{deconst}
The 5D UV completion of the clockwork has  been actively  discussed in the literature \cite{Craig:2017cda, Giudice:2017suc}.
In this section we very briefly mention the possibility to obtain the augmented clockwork potential for the clockworked vev from deconstruction of  5D theories.

The standard clockwork can be generated from a deconstruction of  5d theory of dilatons that can be written as,
\begin{equation}
\label{deconst_2}
\mathcal{S}=2\int_{0}^{\pi R}dy\int d^4x\sqrt{-g}\left(\frac{1}{2}g^{MP}\partial_M\phi\partial_P\phi\right)~,
\end{equation}
and  the 5d metric is given by $ds^2= X(|y|)\left(dx^2+dy^2\right),$ where $y$ is the $\mathbb{S}^1/\mathbb{Z}_2$ orbifolded  extra spatial dimension.
If we consider the  $5^{\textrm{th}}$ dimension $y$ is discretized in $N$ segments by assuming $y_j=ja$, where $j=0,...,N$ and $a=\pi R/N$ is the smallest lattice spacing, on deconstruction and with some field redefinitions \footnote{The relevant field redefinition involved here is $\phi\to\sqrt{2}X^{1/2}Y^{1/4}\phi$.}, Eq.~\eqref{deconst_2} becomes
\begin{equation}
\label{deconst_3}
\mathcal{S}={1\over 2}\int d^4x\left[\sum_{j=0}^{N}(\partial_\mu\phi_j)^2+\sum_{j=0}^{N-1}{N^2X_j\over \pi^2R^2Y_j}\left(\phi_j-{X_j^{1/2}Y_j^{1/4}\over X_{j+1}^{1/2}Y_{j+1}^{1/4}}\phi_{j+1}\right)^2\right]~,
\end{equation}
where we consider $X(|y|)_j =\exp(-4K\pi R j /3N)$ with $K$ as the  clockwork spring constant. This choice gives back the clockwork potential analogous to Eq.~\eqref{clockrev_4}, where we can identify $\Lambda$ and $q$ as follows :
\begin{equation}
\label{deconst_5}
f^{q-1}\Lambda^{3-q}={N^2X_j\over \pi^2R^2Y_j}={N^2\over \pi^2R^2},~~ \textrm{and},~~ q={X_j^{1/2}Y_j^{1/4}\over X_{j+1}^{1/2}Y_{j+1}^{1/4}}=e^{{K\pi R\over N}}.
\end{equation} 

The one-sided clockwork potential of Eq.~\eqref{oneside_1} can be generated by simply adding a potential term at the $y=0$ brane. The 5D action for such case is given by
\begin{equation}
\label{deconst_6}
\mathcal{S}=2\int_{0}^{\pi R}dy\int d^4x\sqrt{-g}\left(\frac{1}{2}g^{MN}\partial_M\phi\partial_N\phi-\delta(y)V_{\textrm{soft}}(\phi)\right)~.
\end{equation}
The generalization of the one-sided clockwork potential to include nearest neighbour interaction (Eq.~\eqref{oneside_11}) is, however, difficult to generate from a 5D deconstructed scenario.

 For the two-sided clockwork, one can chose a slightly different metric for the clockwork spacetime with a special pivot point (say at  $y_k$), as follows
\begin{equation}
\label{deconst_7}
X_j=Y_j=e^{-{4K\pi R\over 3N}|j-k|}~.
\end{equation}
Evidently, for the generation of the two-sided clockwork vev, the breaking term has to be added on the pivot point, \textit{i.e.} at $y=y_k$ brane. Deconstruction using the clockwork metric with above choice followed by some field redefinitions leads to the two-sided clockwork action in 4D as
\begin{eqnarray}
\label{deconst_8}
\nonumber
\mathcal{S}&=& 2\int_{0}^{\pi R}dy\int d^4x\sqrt{-g}\left(\frac{1}{2}g^{MN}\partial_M\phi\partial_N\phi-\delta(y-y_k)V_{\textrm{soft}}(\phi)\right)~,\\
\nonumber\\
&=&{1\over 2}\int d^4x\left[\sum_{j=0}^{N}(\partial_\mu\phi_j)^2+{N^2\over \pi^2R^2}\left(\sum_{j=0}^{k-1}{1\over q^2}\left(q\phi_j-\phi_{j+1}\right)^2+\sum_{j=k}^{N-1}\left(\phi_j-q\phi_{j+1}\right)^2\right)-V_{\textrm{soft}}(\phi_k)\right].~~~~~~
\end{eqnarray}
The clockwork continuum limit ($N\to\infty$) and the UV completions of these 5D models, although interesting to investigate, is however, beyond the mandate of the discussion here.



\begin{center}
\rule{0.5\textwidth}{1pt}
\end{center}
\newpage
 
\bibliographystyle{JHEP}
\bibliography{clockworkvev}


\end{document}